# BigDB: Automatic Machine Learning Optimizer

Anna Pyayt
University of South Florida
pyayt@usf.edu

Michael Gubanov
University of Washington
mgubanov@uw.edu

## ABSTRACT
In this short vision paper, we introduce a machine learning optimizer for data management and describe its architecture and main functionality.

## Categories and Subject Descriptors
H.2.4 [**Database Manager**]

## 1. INTRODUCTION
As Big data agenda recently started gaining momentum [1], the algorithms to turn large amounts of heterogeneous data into easily *accessible* and *actionable* knowledge became important. There are several steps on the way from Big data to structured knowledge and *automatic* actions that can be taken based on this knowledge. First, heterogeneous Big data needs to be automatically classified by data type (e.g. HTML, Video, Audio, Text, Images, etc). Second, depending on the type, specific separately developed Information Extractors (IE) need to be applied to the corpus to extract structured knowledge. Third, a set of important features needs to be defined and for this knowledge. Forth, a set of machine learning classifiers needs to be trained using those features. And, finally, an optimizer needs to be run to automatically iterate over different types of classifiers and their parameters, in order to get the best performance. To the best of our knowledge, even though all the steps described above received significant attention from data management and machine learning research communities [2-9], the last step – optimization of classification performance – is slightly less studied, partially due to the fact that its significance is becoming apparent only on the large scale datasets. Here we focus on the architecture of such an optimizer, intentionally leaving detailed experiments and evaluation for a full paper.

## 2. ARCHITECTURE
Figure 1 illustrates the architecture of the machine learning optimizer. First, the user submits a query to the BigDB engine : "train a classifier using the set of features extracted on the previous step". Second, the optimizer picks a set of different suitable machine learning algorithms and trains them all in parallel using the same features and the same training data from the Big data corpus. Third, after training, the optimizer tests classification performance of all trained classifiers using the test data from the Big data corpus and selects the best performers. Forth, the optimizer changes the parameters of the classification algorithms and/or changes the set of features, retrains the classifiers, tests the performance, selects the best algorithms, and repeats the loop again preset or adjustable number of times to finally output the optimal model.

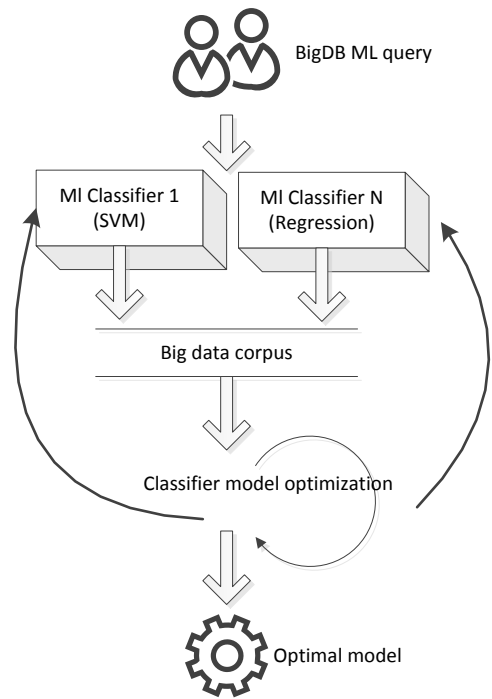

Figure 1. BigDB ML optimizer architecture